\documentclass{PoS}

\title{Multi-Lepton and Isolated Lepton Events at HERA}

\ShortTitle{Multi-Lepton and Isolated Lepton Events at HERA}

\author{\speaker{David M. South}\thanks{on behalf of the H1 and ZEUS Collaborations}\\
        Technische Universit\"{a}t Dortmund\\
        Experimentelle Physik V\\
        44221 Dortmund, Germany\\
        E-mail: \email{david.south@desy.de}}

\abstract{Measurements of the production of events containing isolated
high energy leptons (electrons and muons) produced in $ep$ collisions
have been performed with the H1 and ZEUS detectors at HERA, using data
collected in the period $1994-2007$.
Topologies with more than one charged lepton or with a charged lepton
in coincidence with missing transverse momentum are analysed.
The H1 and ZEUS data, corresponding to an integrated luminosity of
about $0.5$~fb$^{-1}$ per experiment, are combined in a common phase space.
The observed event yields are compared to the predictions from the
Standard Model.
In general a good agreement is found, where the
expectation is dominated by photon--photon collisions for the
multi--lepton topologies and by single $W$ production in the case of
events with an isolated electron or muon and missing transverse
momentum.
Events with large transverse momentum are observed.
Total and differential cross sections of these processes are measured.}

\FullConference{European Physical Society Europhysics Conference on High Energy Physics,
EPS-HEP 2009,\\
July 16 - 22 2009\\
Krakow, Poland}

\begin{document}

\section{Introduction}

The electron--proton collisions at HERA provide a unique opportunity
to look for physics beyond the Standard Model.
Events with one or more isolated leptons in the final state, as well
as in combination with missing transverse momentum, maybe a signature
for rare processes.
The good lepton identification and hadronic final state reconstruction
of the H1 and ZEUS experiments means that such topologies provide a clean signal.
The Standard Model (SM) expectation for such events at HERA is low, so
the analysis benefits from the combination of the H1 and ZEUS data,
resulting in a total integrated luminosity of about $1$~fb$^{-1}$.


\section{Multi-Lepton Events}

The production of multi--lepton final states in $ep$ collisions
proceeds in the SM mainly via photon--photon interactions.
Measurements of multi--lepton production at
HERA have been performed by the H1~\cite{h1multilep}
and ZEUS~\cite{zeusmultilep} collaborations.
A combined analysis of the H1 and ZEUS data is performed in a
common phase space, using the full data samples available to
both experiments~\cite{h1zeusmultilep}.


Electrons are identified in the polar-angle range
$5^\circ < \theta_{e} < 175^\circ$ with $E_{e}>10$~GeV in
the range $5^\circ < \theta_{e} < 150^\circ$ and $E_{e}>5$~GeV
in the backward region $150^\circ < \theta_{e} < 175^\circ$.
Muon candidates are identified in the range
$20^\circ <\theta_{\mu} < 160^\circ$ with $P_{T}^{\mu}>2$~GeV.
All lepton candidates are required to be isolated with respect to each
other by a minimum distance of at least $0.5$ units in the $\eta-\phi$
plane.
At least two central ($20^\circ < \theta < 150^\circ$) lepton candidates
are required, one of which must have $P_T^\ell>10$~GeV and
the other $P_T^{\ell}>5$~GeV. 
Additional leptons identified in the detector according to the
criteria defined above may be present in the event.
According to the number and the flavour of the lepton candidates,
the events are classified into mutually exclusive topologies.
A full description of the common phase space event selection is
presented in~\cite{h1zeusmultilep}.

\begin{figure*}[h]
\centering
\includegraphics[width=0.81\columnwidth]{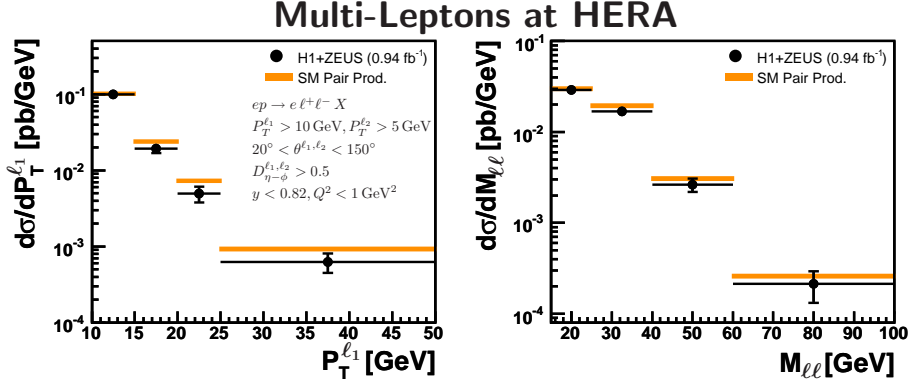}
\vspace{-0.35cm}
\caption{The cross section for lepton--pair photoproduction in a
  restricted phase space as a function of the leading lepton
  transverse momentum $P^{{\ell}_{1}}_{T}$ (a) and the invariant mass
  of the lepton pair $M_{\ell\ell}$ (b). The total error bar is shown,
  representing the statistical and systematic uncertainties added in
  quadrature. The bands represent the uncertainty in the SM prediction.}

\label{fig:mlep}
\end{figure*}

A good overall agreement is observed with the SM in all event
topologies, where the SM prediction in the $e\mu\mu$, $\mu\mu$ and
$e\mu$ topologies is dominated by muon pair production while the
$eee$ and $ee$ topologies contain mainly events from electron pair
production.
%
%
For events where \mbox{$\sum P_T^{\ell}>100$~GeV}, seven events are
observed in the data, compared to $3.13 \pm 0.26$ expected from the
SM, as shown in table~\ref{tab:mlep}.
All seven events were recorded in the $e^+p$ data, for which the SM
expectation is $1.94 \pm 0.17$.
The lepton--pair production cross section is measured in a phase space
dominated by gamma--gamma interactions as
$0.66 \pm 0.03 ({\rm stat.}) \pm 0.03 (\rm sys.)$~pb,
in agreement with the SM prediction of $0.69 \pm 0.02$~pb.
The cross section is also measured as a function of 
$P^{{\ell}_{1}}_{T}$ and the invariant mass of the lepton pair
$M_{\ell\ell}$, as shown in figure~\ref{fig:mlep}.


\begin{table}[h]
\begin{center}
\begin{tabular}{ c c c c c }
Data sample & Data & SM & Lepton Pair Production & Background \\
\hline                                        
e$^{+}$p ($0.56$ fb$^{-1}$) & $7$ & $1.94 \pm 0.17$ & $1.52 \pm 0.14$  & $0.42 \pm 0.07$ \\ 
e$^{-}$p ($0.38$ fb$^{-1}$) & $0$ & $1.19 \pm 0.12$ & $0.90 \pm 0.10$  & $0.29 \pm 0.05$ \\ 
All ($0.94$ fb$^{-1}$) & $7$ & $3.13 \pm 0.26$ & $2.42 \pm 0.21$  & $0.71 \pm 0.10$ \\ 
\hline                                        
\end{tabular}
\end{center}
\vspace{-0.35cm}
\caption{Observed and predicted multi--lepton event yields for $\sum
  P_{T} >$ $100$~GeV in $0.94$~pb$^{-1}$ of HERA data.
  Di--lepton and tri--lepton events are combined.
  The uncertainties on the predictions include model uncertainties
  and experimental systematic uncertainties added in quadrature.}
\label{tab:mlep}
\end{table}


\section{Events with Isolated  Leptons and Missing Transverse Momentum }

In the SM, events containing a high $P_{T}$ isolated electron or muon
and large missing transverse momentum, $P_{T}^{\rm miss}$, arise in
$ep$ collisions from the production of single $W$ bosons with
subsequent decay to leptons.
Events of this topology have been observed at HERA~\cite{h1isolep,zeusisolep}.
A combined analysis of the H1 and ZEUS data has recently been
performed in a common phase space, which makes use of the full data
samples available to both experiments~\cite{h1zeusisolep}.


The event selection is based on those used by H1~\cite{h1isolep}
and ZEUS~\cite{zeusisolep}.
Isolated lepton candidates are required to have $P_{T}^{\ell}>10$~GeV,
to be in the central region of the detector $15^{\circ} < \theta_{\ell} <
120^{\circ}$ and to be isolated from tracks and identified jets in the event.
The event must also exhibit significant missing transverse momentum,
$P_{T}^{\rm miss}>12$~GeV.
Further topological and kinematic cuts are then applied to reject the 
remaining SM background.
A full description of the common phase space event selection is
presented in~\cite{h1zeusisolep}.

\begin{figure*}[h]
\centering
\includegraphics[width=0.5\columnwidth]{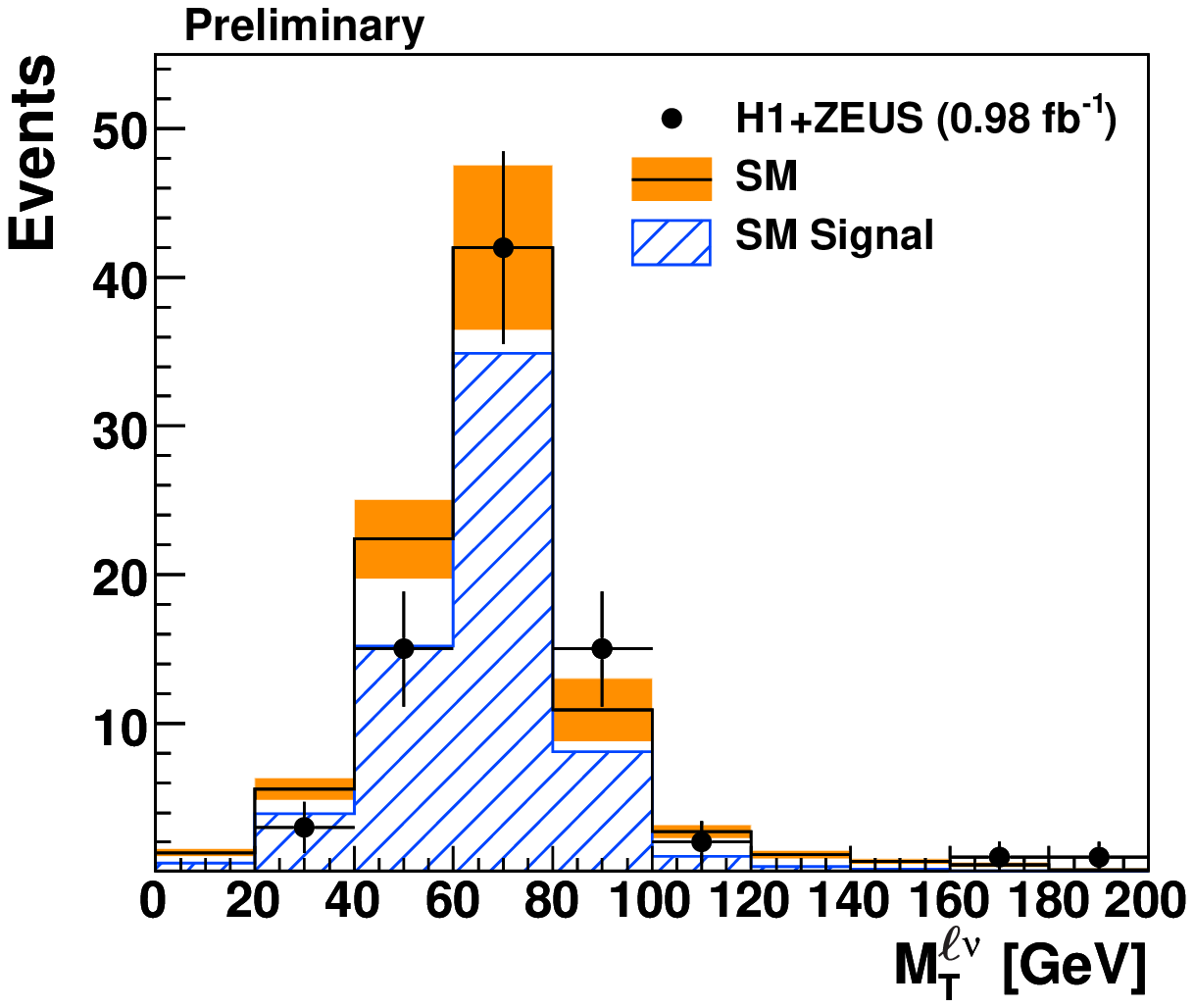}
\includegraphics[width=0.47\columnwidth]{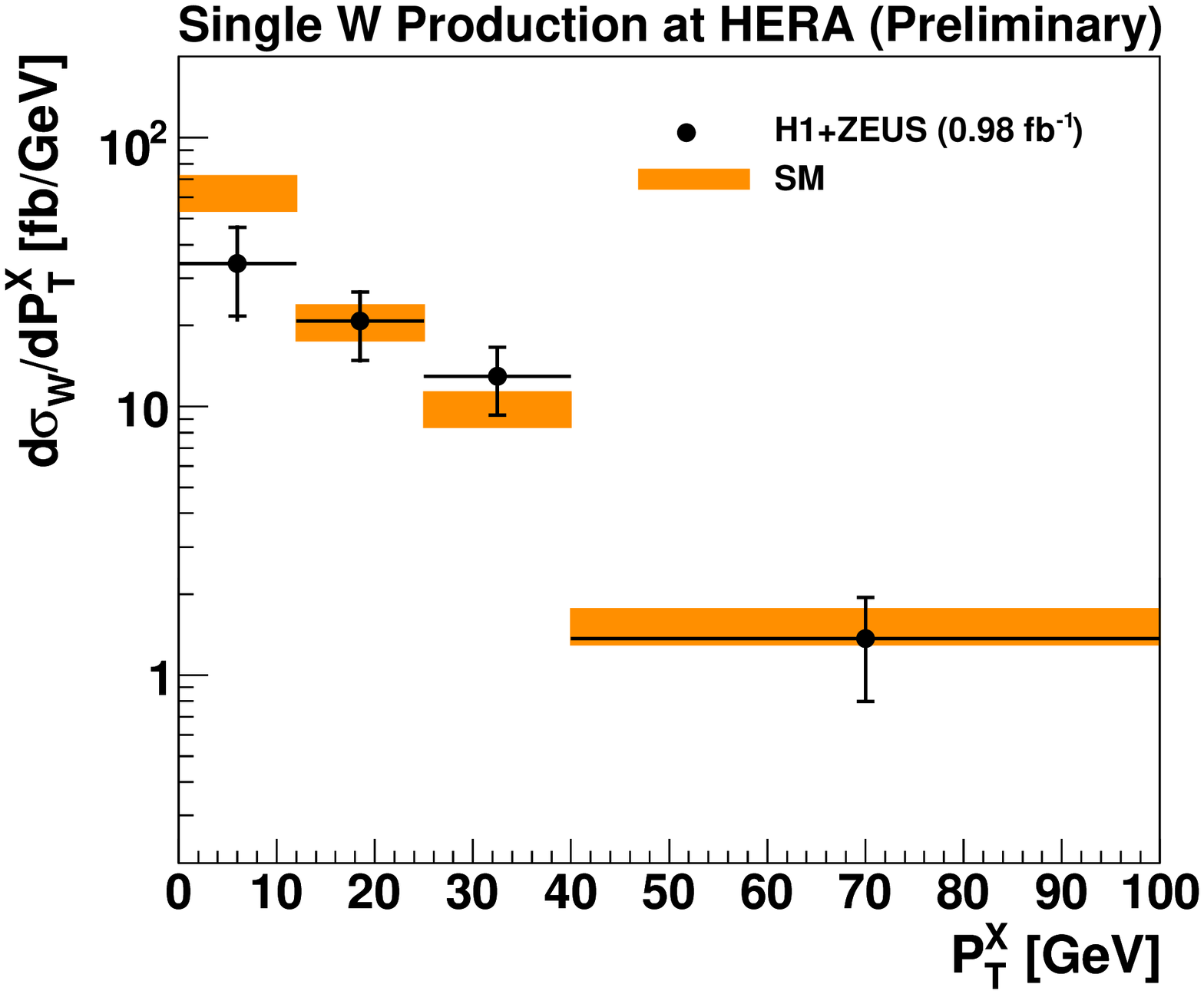}
\vspace{-0.35cm}
\caption{Left: The lepton--neutrino transverse mass $M_{T}^{\ell\nu}$
  of events with an isolated electron or muon and missing transverse
  momentum. The data (points) are compared to the SM
  expectation (open histogram). The signal component of the SM
  expectation, dominated by single $W$ production, is shown as the
  striped histogram. The total uncertainty on the SM expectation is
  shown as the shaded band. Right: The single $W$ production
  cross section as a function of the hadronic transverse momentum,
  $P_{T}^{X}$. The inner error bar represents the statistical error
  and the outer error bar indicates the statistical and
  systematic uncertainties added in quadrature. The shaded band
  represents the uncertainty on the SM prediction.}
\label{fig:isolep}
\end{figure*}

The results of the analysis are shown in table~\ref{tab:isolep}. In
general, a good agreement is observed between the data and the SM
predictions, where the signal component, dominated by single $W$
production, forms the main part of the expectation.
The lepton--neutrino transverse mass distribution is shown in
figure~\ref{fig:isolep}~(left).
For $P_{T}^{X}>25$~GeV, $29$ events are observed in the data, compared
to a SM prediction of $24.0 \pm 3.2$ for the complete HERA $e^{\pm}p$
data.
In the $e^{+}p$ data alone, where a small excess of data is seen in H1
analysis~\cite{h1isolep}, $23$ events are observed in the data,
compared to a SM prediction of $14.0 \pm 1.9$.
The single $W$ cross section is measured as
$1.06 \pm 0.16 ({\rm stat.}) \pm 0.07 (\rm sys.)$~pb,
in agreement with the SM prediction of $1.26 \pm 0.19$~pb.
The cross section is also measured as a function of $P_{T}^{X}$,
as shown in figure~\ref{fig:isolep}~(right).


\begin{table}[h]
\begin{center}
 \begin{tabular}{ c c c c c c }
Channel & & Data & SM & Signal (W) & Background\\
\hline
Electron & Total   & $61$ & $69.2 ~\pm~ ~~8.2$ & $48.3 ~\pm~ 7.4$ & $20.9 ~\pm~ 3.2$ \\
& $P^X_T > 25$ GeV & $16$ & $13.0 ~\pm~ ~~1.7$ & $10.0 ~\pm~ 1.6$ &  ~~$3.1 ~\pm~ 0.7$ \\
\hline
Muon     & Total   & $20$ & $18.6 ~\pm~ ~~2.7$ & $16.4 ~\pm~ 2.6$ &  ~~$2.2 ~\pm~ 0.5$ \\
& $P^X_T > 25$ GeV & $13$ & $11.0 ~\pm~ ~~1.6$ &  ~~$9.8 ~\pm~ 1.6$ &  ~~$1.2 ~\pm~ 0.3$ \\
\hline
Combined & Total   & $81$ & $87.8 ~\pm~ 11.0$  & $64.7 ~\pm~ 9.9$ & $23.1 ~\pm~ 3.3$ \\
& $P^X_T > 25$ GeV & $29$ & $24.0 ~\pm~ ~~3.2$  & $19.7 ~\pm~ 3.1$ &  ~~$4.3 ~\pm~ 0.8$ \\
\hline                                        
\end{tabular}
\end{center}
\vspace{-0.35cm}
\caption{Observed and predicted number of events with an
  isolated electron or muon and missing transverse momentum
  in $0.98$~fb$^{-1}$ of HERA data. The results are shown for the full selected
  sample and for the subsample at large hadronic transverse momentum
  $P_{T}^{X}>25$~GeV.
  The quoted errors contain statistical and systematic
  uncertainties added in quadrature.}
\label{tab:isolep}
\end{table}


\section{Summary}

Analyses of events with multi--leptons and isolated leptons with
$P_{T}^{\rm miss}$ have been recently published individually by H1 and ZEUS.
Combined H1 and ZEUS analyses are also now performed in a common phase
space, to take advantage of the complete HERA high energy data.
In general a good agreement is observed with the SM predictions.
The cross sections of multi--lepton and single $W$ production are
measured with a greater statistical precision.
A few events are observed by both H1 and ZEUS in the $e^{+}p$ data
at high $P_{T}$ and high mass, a region where the SM expectation is low.

\end{document}